\documentclass[aps,prb,citeautoscript,twocolumn,floatfix,showpacs,superscriptaddress]{revtex4}
\usepackage{amsmath,amssymb,graphics}
\begin{document}
\title{Wave functions for quantum Monte Carlo calculations in solids:
  Orbitals from density functional theory with hybrid
  exchange-correlation functionals}

\author{Jind\v rich Koloren\v c}
\altaffiliation[On leave from ]{Institute of Physics, Academy of
  Sciences of the Czech Republic, Na Slovance~2, CZ-18221 Praha~8,
  Czech Republic}
\affiliation{Department of Physics and Center for High Performance
  Simulation, North Carolina State University, Raleigh, North Carolina
  27695, USA}
\affiliation{I. Institut f\"ur Theoretische Physik, Universit\"at Hamburg,
  Jungiusstra\ss e 9, 20355 Hamburg, Germany}
\author{Shuming Hu}
\affiliation{Department of Physics and Center for High Performance
  Simulation, North Carolina State University, Raleigh, North Carolina
  27695, USA}
\author{Lubos Mitas}
\affiliation{Department of Physics and Center for High Performance
  Simulation, North Carolina State University, Raleigh, North Carolina
  27695, USA}

\date{\today}

\begin{abstract}
We investigate how the fixed-node diffusion Monte Carlo energy of
solids depends 
on single-particle orbitals used in Slater--Jastrow wave
functions. We demonstrate that
the dependence can be significant, in particular in the case of 3$d$
transition-metal compounds, which we adopt as examples. 
We illustrate how exchange-correlation functionals with variable
exact-exchange component 
can be exploited to reduce the fixed-node errors. On the basis of
these results we argue that the fixed-node quantum Monte Carlo
provides a variational approach for optimization of effective
hamiltonians with parameters.
\end{abstract}









\maketitle

\section{Introduction}\label{sec:intro}

Steady increase in supercomputer performance  
over the last three decades stimulates development of very accurate
electronic structure methodologies that provide detailed understanding
of ever larger and more complex systems. Diffusion Monte Carlo (DMC)
method\cite{foulkes2001,hammond1994} is one of the prime examples
of these advanced many-body approaches. It uses a stochastic process 
to refine a given many-body wave function $|\Psi_T\rangle$ towards the actual
ground-state solution $|\Psi_0\rangle$. The antisymmetry of the wave
function with respect to fermionic degrees of freedom is usually
maintained by imposing the so-called fixed-node condition. In practice
it means that the nodal surface (the subset of configuration space
where the many-body wave function vanishes) is restricted to be the
same as in the initial guess $|\Psi_T\rangle$ throughout the entire stochastic
simulation. The basic premise is that high accuracy is achieved even
when relatively simple functional forms are employed for
$|\Psi_T\rangle$.

The impact of the fixed-node approximation has been intensively
studied in few-electron atoms and molecules as well as in homogeneous
systems, where the performance of the DMC method with increasing
accuracy of the trial wave function $|\Psi_T\rangle$ is relatively
well mapped out\cite{casula2004,bajdich2006,lopezrios2006,umrigar2007,
kwon1998,holzmann2006}. Applications to solids are not nearly as
numerous, and hence the influence of the fixed-node condition in
crystalline settings is much less examined. Evaluation of bulk
properties necessarily involves extrapolation to the thermodynamic
limit, which decreases the amount of computational resources available
for exploration of rather subtle fixed-node errors. Consequently, only
the simplest trial wave functions, having the Slater--Jastrow form,
are typically employed. These wave functions have nodal
surfaces defined by single-particle orbitals, and therefore they
correspond to mean-field nodes. The subject of our study are
the fixed-node errors associated with such trial functions when
they are applied to simple transition-metal oxides in crystalline phases.

\section{Fixed-node diffusion Monte Carlo}\label{sec:DMC}

The fixed-node diffusion Monte Carlo method calculates expectation
values of quantum-mechanical operators in the ground state (as well as
in certain excited states) of a many-body hamiltonian $\hat H$. The 
DMC wave function is found by a projection
\begin{subequations}
\begin{equation}\label{eq:DMC_proj}
|\Psi_D\rangle=\lim_{\tau\to\infty} e^{-\tau\hat H} |\Psi_{T}\rangle
\end{equation}
that gradually increases the weight of the lowest energy eigenstate of
$\hat H$ relative to all other states contained in some initially
guessed wave function $|\Psi_{T}\rangle$. Building on similarity
between the Schr\" odinger and the diffusion equations, the projection
is realized with the aid of a classical stochastic process. The
outcome of this simulation is a set of $3N$-dimensional samples
$\{R_i\}$ distributed according to a probability distribution
$\mathcal{P}(R)=\Psi_D(R)\Psi_{T}(R)/\langle\Psi_D|\Psi_{T}\rangle$.
Here $R=(\mathbf{r}_1,\mathbf{r}_2,\dots, \mathbf{r}_N)$ denotes
coordinates of all $N$ electrons comprising the investigated quantum
system. The hamiltonian $\hat H$ is assumed to be spin independent,
which prevents any spin-flip processes to occur in
Eq.~\eqref{eq:DMC_proj}, and hence the spins do not enter the
simulations as dynamical variables. The probabilistic interpretation
of $\mathcal{P}(R)$ is possible only if it is a positive quantity. In
the case of fermions, the projection as written in
Eq.~\eqref{eq:DMC_proj} does not fulfill this requirement,
which must therefore be prescribed in the form of an additional
condition
\begin{equation}
\Psi_D(R) \Psi_{T}(R) \geq 0\,.
\end{equation}
\end{subequations}
This step introduces the so-called fixed-node approximation, since the
projection cannot reach the true ground state $|\Psi_0\rangle$ if
the fermionic nodes of the trial wave function $|\Psi_T\rangle$ differ
from (a priori unknown) nodes of the desired ground state. Our primary
aim is
to explore the impact of this fixed-node
approximation for a particular functional form of the trial
wave function.

The set of $\mathcal{N}$ samples $\{R_i\}$ acquired as a result of a
DMC simulation 
can be directly used to calculate the so-called mixed estimates of the
quantum-mechanical expectation values,
\begin{align}\label{eq:mix_estim}
\frac{\langle\Psi_D|\hat A|\Psi_{T}\rangle}{%
   \langle\Psi_D|\Psi_{T}\rangle}
&=\int dR \, \biggl(\frac{\Psi_D(R)\Psi_{T}(R)}{%
   \langle\Psi_D|\Psi_{T}\rangle}\biggr)
   \biggl(\frac{\bigl[\hat A\Psi_{T}\bigr](R)}{\Psi_{T}(R)}\biggr)
   \nonumber\\
&=\frac1{\mathcal{N}} \sum_{i=1}^{\mathcal{N}}
   \frac{\bigl[\hat A\Psi_{T}\bigr](R_i)}{\Psi_{T}(R_i)}
   +\mathcal{O}(\mathcal{N}^{-1/2})\,.
\end{align}
If the operator $\hat A$ commutes with the hamiltonian, the mixed
estimate equals the desired quantum-mechanical expectation value,
$\langle\Psi_D|\hat A|\Psi_{T}\rangle/
\langle\Psi_D|\Psi_{T}\rangle
= \langle\Psi_D|\hat A|\Psi_D\rangle/
\langle\Psi_D|\Psi_D\rangle$.
In general, however, there is an error proportional to
the difference between $|\Psi_D\rangle$ and $|\Psi_{T}\rangle$
that can be reduced to the next order using  
the following extrapolation \cite{hammond1994}
\begin{multline}\label{eq:extrap_estim}
\frac{\langle\Psi_D|\hat A|\Psi_D\rangle}{%
\langle\Psi_D|\Psi_D\rangle}
 = 2\, \frac{\langle\Psi_D|\hat A|\Psi_{T}\rangle}{%
    \langle\Psi_D|\Psi_{T}\rangle}
 - \frac{\langle\Psi_{T}|\hat A|\Psi_{T}\rangle}{%
    \langle\Psi_{T}|\Psi_{T}\rangle}   \\
   +\mathcal{O}\biggl(\biggl|
      \frac{\Psi_D}{\sqrt{\langle\Psi_D|\Psi_D\rangle}}
      -\frac{\Psi_{T}}{\sqrt{\langle\Psi_{T}|\Psi_{T}\rangle}}
   \biggr|^2\biggr).
\end{multline}
The expectation value
$\langle\Psi_{T}|\hat A|\Psi_{T}\rangle/\langle\Psi_{T}|\Psi_{T}\rangle$
is evaluated by 
straightforward Monte Carlo integration and it is referred to as
the estimate of the variational Monte Carlo (VMC) method.

\section{Trial wave functions}\label{sec:psiT}

We employ trial wave functions having the Slater--Jastrow functional
form, which is an antisymmetrized product of single-particle orbitals
(Slater determinant) multiplied by a correlation factor that is
symmetric with respect to pair-electron exchanges. We can write
\begin{equation}\label{eq:wf_sj}
\Psi_T(R)=\mathop{\rm det}\{\psi^{\uparrow}_{\alpha}\}
          \mathop{\rm det}\{\psi^{\downarrow}_{\beta}\}\,
          e^{J(R,X)}\,,
\end{equation}
where $\psi^{\uparrow}_{\alpha}$ and $\psi^{\downarrow}_{\beta}$ are
spatial parts of single-particle orbitals that correspond to spin-up
respectively spin-down electronic states. The vector
$X=(\mathbf{x}_1,\mathbf{x}_2,\dots, \mathbf{x}_M)$ encompasses
positions of all $M$ ions in the lattice.  The expression in
Eq.~\eqref{eq:wf_sj} represents only a spatial component of the trial
wave function corresponding to one particular spin configuration,
where electrons with labels $1,\dots,N_{\uparrow}$ are in the spin-up
state and electrons with labels $N_{\uparrow}+1,\dots,N$ are in the
spin-down state. We can use this simplified form with fixed spin
states instead 
of the full wave function as long as neither the hamiltonian nor any
other operator, expectation value of which we intend to calculate,
depend on electron spins. The applications we consider involve only
cases with zero total spin, i.e., $N$ is an even number and
$N_{\uparrow}=N_{\downarrow}=N/2$.

The Jastrow correlation factor we use,
\begin{equation}\label{eq:wf_j}
J(R,X)= \sum_{i,j} f(\mathbf{r}_i-\mathbf{r}_j)
      + \sum_{i,I} g(\mathbf{r}_i-\mathbf{x}_{I})\,,
\end{equation}
contains one- and two-body terms, $g$ and $f$, that are parametrized
in the same way as in Ref.~~\onlinecite{wagner2007a}. This correlation
factor improves efficiency of the Monte Carlo sampling and accuracy of
general expectation values calculated according to
Eq.~\eqref{eq:extrap_estim}. Quality of the DMC total energy
$E_D$ depends solely on the accuracy of the
nodal surface that is, given the functional form of
Eq.~\eqref{eq:wf_sj}, completely determined by the single-particle
orbitals $\{\psi^{\uparrow}_{\alpha},\psi^{\downarrow}_{\beta}\}$.
Ideally, these orbitals would be parametrized by an expansion in a saturated
 basis with the expansion coefficients varied to minimize the
DMC total energy. Unfortunately, the stochastic noise and 
the computational demands of the DMC method make
this route extremely inefficient in practice except,
perhaps, in the case of the simplest few electron systems.

A more feasible alternative is to skip the DMC projection,
Eq.~\eqref{eq:DMC_proj}, and optimize the orbitals
$\{\psi^{\uparrow}_{\alpha},\psi^{\downarrow}_{\beta}\}$ with respect
to a simpler quantity than $E_D$. For instance, the variational
Monte Carlo methodology can be employed to minimize the variational
energy $E_V=\langle\Psi_{T}|\hat H|\Psi_{T}\rangle /
\langle\Psi_{T}|\Psi_{T}\rangle$. 
The VMC optimization of one-particle orbitals was successfully employed for
atoms and small molecules of the first-row
atoms\cite{filippi1996,umrigar2007,toulouse2008}, but the method is
still too demanding for applications to solids.  It is also not
completely robust, since an improvement of the variational energy
does not automatically guarantee an improvement of the fixed-node
energy due to the limited parametric freedom of the trial wave
function $|\Psi_{T}\rangle$.
This issue is even more pronounced when trial wave functions are not
optimized with respect to the VMC energy, but with respect to another
quantity, such as the energy variance.

To avoid the large number of variational parameters needed to describe
the single-particle orbitals, another family of methods has been
proposed. The orbitals in the Slater--Jastrow wave function are found
as solutions to self-consistent-field (SCF) equations that represent a
generalization of the Hartree--Fock
equations.\cite{filippi2000,umezawa2003}
The correlations described by the Jastrow factor enter the SCF
equations via variational Monte Carlo techniques. These methods were
tested in atoms as well as in solids\cite{filippi2000,sakuma2006},
albeit only rather limited variational freedom was allowed in the
employed Jastrow factors. Unfortunately, wave functions derived in
this way did not lead to lower DMC energies compared to wave functions
with orbitals from the Hartree--Fock theory or from the local density
approximation (LDA).\cite{filippi2000,prasad2007} It is unclear,
whether the lack of observed improvements in the fermionic nodal surfaces
stems from insufficient flexibility of the employed correlation
factors or from too small set of tested cases, all with only $s$ and
$p$ valence electrons.

In this paper we also use SCF equations as a means to construct the
one-particle orbitals, but the parametric dependence of these
equations is introduced without any relation to the actual Jastrow
factor. The self-consistent-field equations in question are Kohn--Sham
equations corresponding to the hybrid exchange-correlation functional
PBE0${}_w$,\cite{perdew1996b}
\begin{equation}
\label{eq:pbe0}
E^{PBE0_w}_{xc}=w E_x^{HF} + (1-w) E^{PBE}_{x} + E^{PBE}_c\,.
\end{equation}
Here $E^{PBE}_{x}$ and $E^{PBE}_c$ are exchange and correlation parts
of the PBE form\cite{perdew1996} of the generalized gradient
approximation (GGA), and $E_x^{HF}$ is the exchange from the Hartree--Fock
theory. The weight $w$ is in the range $0<w<1$ and serves as a
variational parameter with respect to which the fixed-node DMC energy
is minimized.

In the transition-metal oxides we study it is understood
that exchange in GGA is underestimated,\cite{anisimov1991} whereas
the ``exact'' exchange from the Hartree--Fock theory overestimates the
real exchange mechanism as any screening effects are
neglected.\cite{towler1994,needs2003} The hybrid density-functional
theory (DFT) provides an interpolating scheme between these two
extremes. There are numerous examples in the literature illustrating
that reasonable agreement of its predictions and experimental
observations can be achieved.\cite{alfredsson2004,franchini2005,
kolorenc2007,kummel2008}

Arguably, the single parameter introduced to the one-particle
orbitals represents only relatively constrained variational freedom
compared to the approaches mentioned
above.\cite{filippi1996,umrigar2007,toulouse2008} On the other hand,
the simplicity of the parameter space allows for direct optimization
of the fixed-node DMC total energy, which improves robustness compared
to minimization within the VMC method. Initial applications of this
strategy to molecules\cite{wagner2003,wagner2007a} and in a more
elementary form also to solids\cite{sola2009} were reported
previously. In the following sections we perform the DMC optimization
for two compounds, MnO and FeO, and systematically analyze the
variational freedom available within this method in crystalline
environments.

\section{Thermodynamic limit}
\label{sec:fse}

Crystals in our simulations are represented by periodically repeated
simulation cells of a finite size, in which the Coulomb interaction
energy is evaluated with the aid of the Ewald
formula.\cite{allen1989,foulkes2001} To calculate bulk quantities
one has to perform an extrapolation to the thermodynamic
limit (infinite crystal volume). Our main
objective, however, is to compare energies obtained for different
trial wave functions in a given simulation cell with {\itshape fixed}
volume, and hence the size extrapolation seems redundant. Indeed, the
leading-order term of the finite-size errors related to the long-range
character of the Coulomb interaction is a function of the average
charge density alone.\cite{chiesa2006,drummond2008} Variation of
single-particle orbitals does not change this quantity, and these
finite-size errors are therefore mostly irrelevant for our wave function
optimization, since they cancel out.

The Coulomb interaction is not the only cause of finite-size biases in
calculations performed in finite simulation cells. There is another
source of non-vanishing surface terms that appear even if no two-body
interaction is present in the hamiltonian.\cite{lin2001} These can be
brought to light as follows: the statement ``periodically repeated
simulation cells'' used above refers to observable quantities and does not
fully specify the boundary condition for the phase of the wave
function. In the language of single-particle description of solids,
different boundary conditions compatible with periodicity of
observables correspond to simulations being done at different $k$
points. Averaging over all $k$ points from the first Brillouin zone
removes the dependence of the computed quantities on the boundary
conditions and in the case of non-interacting particles it is
equivalent to performing the thermodynamic limit. For hamiltonians
with particle-particle interactions this correspondence is not exact,
but a very substantial reduction of finite-size errors is 
observed nevertheless.\cite{lin2001}

Comparison of trial wave functions at a single $k$ point is certainly
a valid approach. On the other hand, variation of the
exchange-correlation functional in the Kohn--Sham equations modifies
the resulting band structure in a generally non-trivial $k$-dependent
manner, see for instance Ref.~~\onlinecite{alfredsson2004} for
illustration. As a result, the minimization of the total energy calculated
at a single $k$ point does not necessarily lead to the same optimal
weight $w$ as does the minimization of the $k$-averaged total
energy. Since the latter is arguably a better approximation of the
thermodynamic limit, which is what we are ultimately interested in, we
work with the $k$-averaged quantities. 

\section{M\lowercase{n}O at ambient conditions}
\label{sec:MnO}

\begin{figure}
\resizebox{\linewidth}{!}{\includegraphics{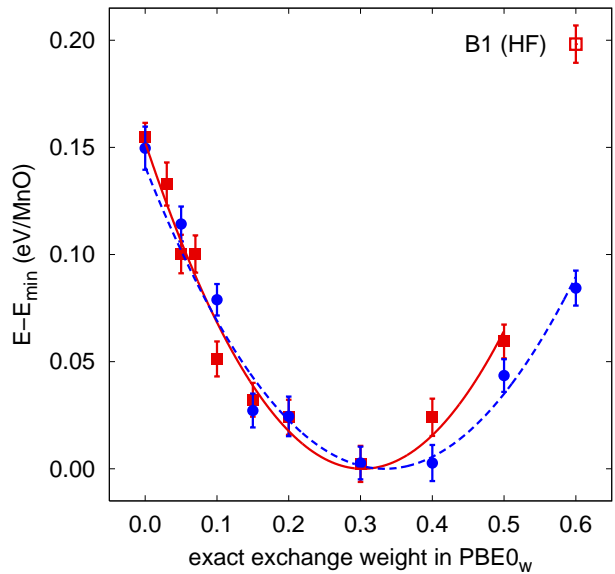}}
\caption{\label{fig:mno_b1b8}(color online) The DMC total energy of MnO
  at experimental equilibrium volume ($V=21.7$ \AA${}^3$/MnO) for two
  distinct structural phases: B1 AFM-II (red squares, solid line) and
  B8 AFM (blue points, dashed line). Minima of the fits are chosen as
  reference energies for the respective phases. Also 
  shown is the DMC total energy of the B1 AFM-II phase
  obtained with the Hartree--Fock orbitals (empty red square).}
\end{figure}

\begin{table}
\caption{\label{tab:optimal_weight}Exact-exchange weight $w$ leading
  to minimal DMC total energies of several phases of MnO and FeO solids.}
\begin{tabular}{llcc}
\hline\hline
compound  & phase & $V$ (\AA${}^3$/$X$O) & $w$\\
\hline
MnO & B1 AFM-II & 21.7 & $0.30\pm 0.01$   \\
MnO & B8 AFM    & 21.7 & $0.33\pm 0.01$   \\
\hline 
FeO & B1 AFM-II & 20.4 & $0.25\pm 0.02$   \\
FeO & B1 AFM-II & 17.3 & $0.217\pm 0.006$ \\
FeO & iB8 AFM   & 17.0 & $0.156\pm 0.003$ \\
\hline\hline
\end{tabular}
\end{table}

We start our investigations with manganese monoxide at experimental
equilibrium volume, $V=21.7$ \AA${}^3$/MnO. Calculations are performed
for two lattice structures, both with antiferromagnetic (AFM) ordering
of magnetic Mn atoms: B1 (symmetry group R$\bar{3}$m, the so-called
AFM-II state) and B8 (symmetry group P$\bar{3}$m1). The former
phase is the low-temperature ground state structure
at atmospheric pressure and the latter is a phase stable at high
pressures.\cite{yoo2005} Since all states investigated in this paper
are antiferromagnetic, we will often leave out this attribute.

Data were collected in simulation cells containing 16 atoms (8 Mn and
8 O). Atomic cores were replaced by norm-conserving pseudopotentials
within the so-called localization approximation.\cite{mitas1991}
Helium core was excluded from oxygen atoms\cite{ovcharenko2001} and
neon core from manganese atoms\cite{lee_private}, which left 168
valence and semi-core electrons explicitly included in the
simulations.

The Monte Carlo simulations were performed with
\textsc{QWalk}\cite{wagner2009} and the single-particle orbitals were
prepared in \textsc{Crystal2003}.\cite{crystal2003} In both codes, the
orbitals were expanded in a gaussian basis, completeness of which
was verified against a converged basis of linearized augmented plane
waves as implemented in the \textsc{Wien2k} code\cite{wien2k}. The
basis set benchmarks were performed within DFT using the PBE-GGA
exchange-correlation functional.

The acquired dependence of the DMC total energies on the
exact-exchange weight $w$ is plotted in
Fig.~\ref{fig:mno_b1b8}. Each energy value shown in the picture
represents an average over 8 $k$ points. Such average made within
DFT differs from the converged integral over the first Brillouin zone
by less than the statistical error bars plotted in the figure. The
quadratic functions fitted through the data using the least-squares
method lead to the optimal values for the weight $w$ listed in
Tab.~\ref{tab:optimal_weight}.  The behavior is virtually the same in
both phases and the optimal weight is very close to the standard PBE0
choice\cite{perdew1996b} $w=0.25$. Drop of the total energy from the
pure PBE-GGA orbitals ($\Leftrightarrow$ PBE0${}_{w=0}$) to the
minimum is $\approx 0.15$ eV/MnO, and from the Hartree--Fock orbitals
($\not\Leftrightarrow$ PBE0${}_{w=1}$) it is $\approx 0.2$ eV/MnO. The
dependence of the fixed-node DMC total energy on the admixture of
the exact exchange is weaker in the MnO solid than in the MnO molecule,
where the gain from the Hartree--Fock to hybrid DFT orbitals was observed
as large as $\approx 0.6$ eV.\cite{wagner2007a} We attribute this
difference to a greater freedom of charge density to adapt in the
molecule than in the constrained solid state environment.

The data collected so far can be utilized to evaluate quantities of
direct physical interest. The difference between the optimized total
energies provides a measure of relative stability of the investigated
phases at the given volume. We find that the
B1 structure is lower in energy than the B8 structure by $0.27\pm
0.01$ eV/MnO. The cohesive energy $E_{coh}$ of MnO crystal can also be
estimated. To this end, energies of isolated Mn and O atoms need to be
calculated, and the crystalline total energy has to be extrapolated to
the infinite volume. We provide details of these steps in
Appendix~\ref{sec:MnO_cohesion}. In the end we obtain $E_{coh}=9.29\pm
0.04$ eV that is in good agreement with the value $9.5$ eV derived
from experimental formation enthalpies.\cite{crc} Our result
differs from the earlier DMC estimate\cite{lee2004,mitas2010}, which is not
surprising as the present approach involves more advanced optimization
of the trial wave function as well as much more thorough analysis of
the finite-size errors.

\section{Are large simulation cells necessary?}
\label{sec:MnO_1u}

\begin{figure}
\resizebox{\linewidth}{!}{%
\includegraphics{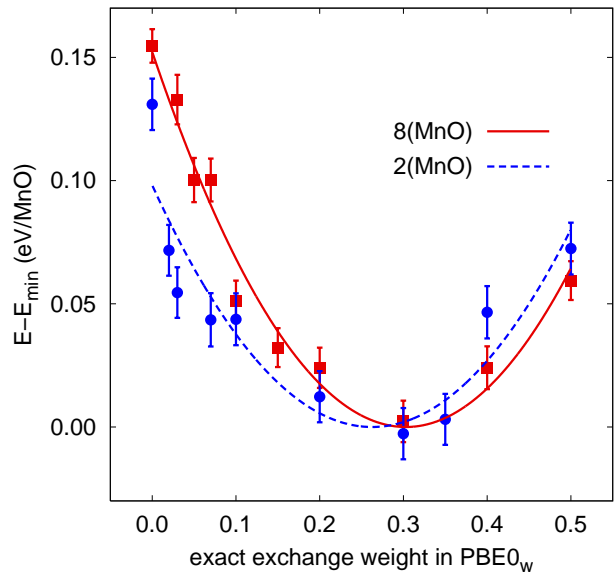}}
\caption{\label{fig:mno_1u4u}(color online) The DMC total energy of MnO 
  (B1 AFM-II phase at $V=21.7$ \AA${}^3$/MnO) as a
  function of exact-exchange weight $w$. Compared are data obtained in
  simulation cells of two different sizes: 16 atoms (red squares, solid
  line) and 4 atoms (blue circles, dashed line). The lines are
  quadratic fits. Minima of these fits are chosen as
  reference energies for the respective data sets.}
\end{figure}

It was argued in Sec.~\ref{sec:fse} that $k$ point averaging should
remove nearly all finite-size biases relevant to the optimization of
single-particle orbitals performed at a fixed volume. It would
certainly be beneficial if one could find the optimal exact-exchange
weight in a small simulation cell and only then proceed with
production runs in large cells.

We have repeated the optimization procedure in
the primitive cell of the B1 AFM-II structure, which contains only 4
atoms (2 Mn and 2 O). All other parameters were kept unchanged, except
the number of $k$ points employed in the averaging had to be
substantially increased---from 8 to 125---to achieve comparable
convergence.%
\footnote{The 125 $k$ points sample the whole Brillouin
zone, they fold into 19 in the irreducible wedge of the Brillouin
zone.} 
The necessity to enlarge the set of considered $k$ points introduces
an extra technical complication, since trial wave functions
corresponding to the majority of these $k$ points are
complex-valued. Consequently, we replace the fixed-node
DMC method with its natural generalization, the so-called fixed-phase
DMC\cite{ortiz1993}. The fixed-phase condition reduces to
the fixed-node condition for real-valued trial wave functions.

The $w$-dependence of the DMC total energy calculated in the
small simulation cell (Fig.~\ref{fig:mno_1u4u}) is similar but not
completely identical to the behavior observed in the larger cell. The
minimum is shifted to a slightly lower weight $w$ and the $E(w)$ curve
raises a little slower with $w$ decreasing towards the pure
PBE-GGA. It is plausible to assume that these differences are a
fingerprint of the residual Coulomb finite-size effects. 

\section{Compressed F\lowercase{e}O}
\label{sec:FeO_pressure}

\begin{figure}
\resizebox{\linewidth}{!}{\includegraphics{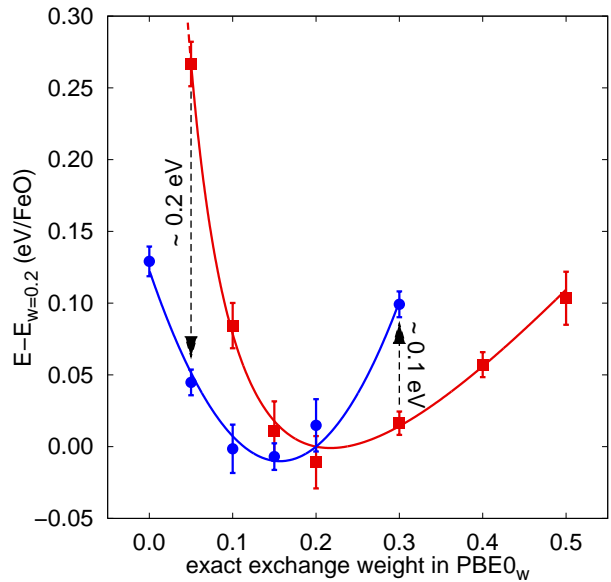}}
\caption{\label{fig:feo_b1ib8}(color online) The DMC total energy of
  compressed FeO in two phases: B1 AFM-II at $V=17.3$ \AA${}^3$/FeO
  (red squares), and iB8 AFM at $V=17.0$ \AA${}^3$/FeO (blue
  points). Lines are least square fits with a quadratic function in
  the iB8 phase, and with $E(w)=(a+bw+cw^2)/(d+w)$ in
  the B1 structure. Note the different choice of reference energies
  compared to Figs.~\ref{fig:mno_b1b8} and~\ref{fig:mno_1u4u}.
  Here the zero energy for each phase is the value of the
  corresponding fit at $w=0.2$.}
\end{figure}
 
Finally, we turn our attention to iron oxide subject to high pressures
at low temperatures. We revisit our earlier study of the phase
transition from the B1 AFM-II phase, stable at atmospheric pressure,
to the iB8 AFM structure (space group P$\bar{6}$m2), which the
fixed-node DMC has predicted as stable above approximately 65
GPa.\cite{kolorenc2008} Calculations leading to this estimate used
single-particle orbitals provided by the PBE0${}_{w=0.2}$
functional. This choice was based on prior
investigations\cite{wagner2003,wagner2007a} and on a preliminary version
of the results we present here. In the following, we analyze in detail
how appropriate the PBE0${}_{w=0.2}$ orbitals are in this case and how
sensitive the transition pressure is to variations of single-particle
orbitals in the Slater--Jastrow trial wave function.

We return back to simulation cells containing 16 atoms, partly because
the orbital optimization appears more robust in larger cells 
(Sec.~\ref{sec:MnO_1u}) and partly because we already acquired
some data in the 16 atom cells in the course of our earlier
investigations. The optimized values for the exact-exchange weight $w$ in
FeO are listed in Tab.~\ref{tab:optimal_weight} for experimental
equilibrium volume $V=20.4$ \AA${}^3$/FeO as well as for two
compressed states. The optimal proportion of the exact exchange
decreases with compression, which is an expected phenomenon---the role
of screening increases as the bands widen and a larger fraction of $d$
electrons participates in chemical bonding.

In contrast to MnO, where B1 and B8 phases displayed very similar
behavior, the two investigated structures of FeO differ noticeably at
comparable volumes. Detailed data for compressed FeO are shown in
Fig.~\ref{fig:feo_b1ib8} to highlight the differences. Like
in Sec.~\ref{sec:MnO}, each total energy is obtained as an average
over 8 $k$ points. To facilitate comparison with our previous
study\cite{kolorenc2008}, energies corresponding to the
PBE0${}_{w=0.2}$ orbitals are used as a reference. We can see that
$w=0.2$ indeed represents a reasonable compromise value, since
corresponding DMC energies lie within error bars from the true minima.

In the case of the iB8 phase, the $E(w)$ data are well described by a
quadratic function. In the B1 phase, on the other hand, the functional
dependence is asymmetric around the minimum and a quadratic function
does not provide a satisfactory fit. To locate the minimum for
Tab.~\ref{tab:optimal_weight}, we used an alternative fitting function
$E(w)=(a+bw+cw^2)/(d+w)$, which characterizes the calculated energies
much better. Figure~\ref{fig:feo_b1ib8} shows that the energy raises
rather fast when the exchange-correlation functional approaches the pure
PBE-GGA. It does not come as a surprise, since the Kohn--Sham spectrum
is metallic in the limit $w\to 0$, which is at odds with
experimental facts. All other
phases investigated here (FeO iB8 and both MnO phases) are insulating
for all values of the weight $w$, and hence even orbitals close to
the PBE-GGA provide reasonable trial wave functions.

The different behavior of the DMC total energies in the B1 and iB8
structures causes the pressure of the transition between these two
phases to depend on the used orbitals. To roughly estimate the
variation of the transition pressure, we assume that the energy--volume
equations of state of the respective phases only rigidly shift along the
energy axis when the exchange weight is varied. The equations of state
corresponding to the PBE0${}_{w=0.2}$ orbitals were calculated in
Ref.~~\onlinecite{kolorenc2008} and the corresponding shifts can be
extracted from Fig.~\ref{fig:feo_b1ib8}. When, for instance, the
PBE0${}_{w=0.3}$ orbitals are used, the iB8 phase is raised in energy
by approximately $0.1$ eV compared to the B1 structure, which leads to
the transition pressure increased to $\approx 85$ GPa. When the
PBE0${}_{w=0.05}$ orbitals are utilized, the iB8 phase is lowered by
approximately $0.2$ eV, which corresponds to the transition pressure
of only $\approx 30$ GPa. Evidently, the B1 to iB8 transition pressure
is quite sensitive to the choice of the single-particle orbitals. Of
course, the DMC method provides a definite prediction as long as the
exchange weight is individually optimized in each phase. The pressure
$65\pm 5$ GPa derived in Ref.~~\onlinecite{kolorenc2008} remains valid
as the actual DMC estimate for the B1 to iB8 transition pressure,
since the energies obtained with the PBE0${}_{w=0.2}$ orbitals lie
within error bars from the minimal energies
(Fig.~\ref{fig:feo_b1ib8}).

\section{Optimization of effective hamiltonians}
\label{sec:oef}

\begin{figure}
\resizebox{\linewidth}{!}{\includegraphics{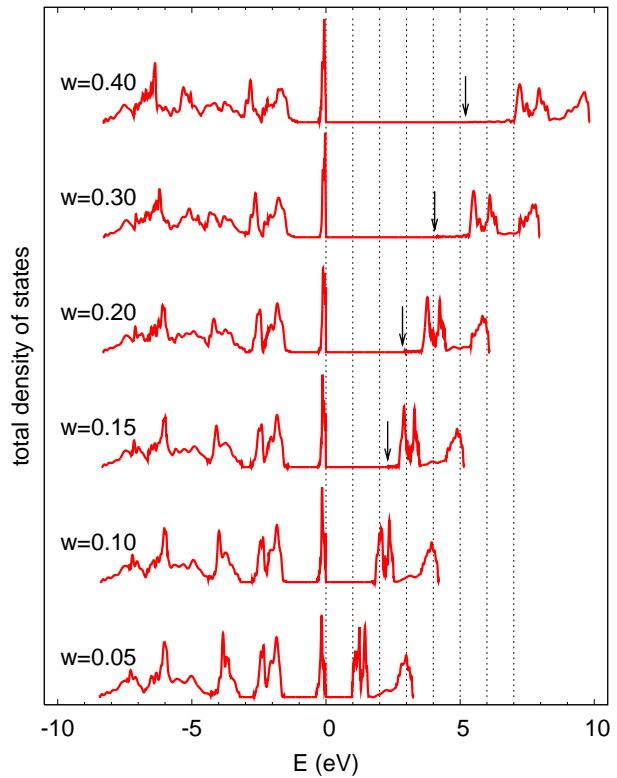}}
\caption{\label{fig:feo_b1dos}(color online) Density of states of FeO
  (B1 phase at $V=20.4$ \AA${}^3$/FeO) from PBE0$_w$ calculations with varying 
$w$. Where it is not obvious, the bottom of the conduction band is
indicated with an arrow.}
\end{figure}

The optimization of not only the variational wave function but also of
the effective hamiltonian can be considered in a broader context. The
upper-bound property of the fixed-node approximation offers a new
opportunity for finding optimal values of {\em any parameters} of
effective hamiltonians in a consistent and well-defined manner. For
example in DFT+$U$ methods,\cite{anisimov1991} the Hubbard parameter
$U$ as well as the form of the double counting terms could be optimized
in a similar way as the exact-exchange weight in the 
presented calculations. The optimized hamiltonian can
subsequently be utilized for further calculations. We illustrate this
on the density of states of the iron oxide shown as a function of the
exact-exchange weight in Fig.~\ref{fig:feo_b1dos}. Note that for
PBE0$_{w=0.2}$, i.e., for the functional that we used as optimal for the
calculations of the equation of state, the single-particle spectrum
exhibits a reasonable value of the gap of about 2.9 eV, which is much
closer to the experimental value ($\approx 2.4$ eV,
Ref.~~\onlinecite{bowen1975}) than either pure GGA or pure 
Hartree--Fock limits. The key point is that although the
hamiltonian was optimized in the ground state, the excitations are
clearly improved as well.

In a more general sense, one could consider effective
hamiltonians with more parameters and/or with more elaborated content
beyond the one-particle form, such as explicit treatment of particle
pairs, for example. The idea of employing the
fixed-node DMC method could be of importance for such constructions as
the most accurate and explicitly variational method available at present.

\section{Conclusions}

We have found that single-particle orbitals in Slater--Jastrow wave
functions represent a non-trivial variational parameters for fermionic
nodes in $3d$ transition-metal compounds. When these orbitals are
generated with the aid of an exchange-correlation functional with
variable admixture of the exact exchange, the corresponding fixed-node
DMC energies differ by several tenths of eV per transition-metal
atom. These variations can translate to
substantial quantitative changes in the phase diagram as demonstrated
in the case of iron oxide. Consequently, some form of orbital
optimization should be performed in order to confidently predict
relative stability of different crystal structures and the location
of corresponding phase transitions.

The optimal amount of exact exchange providing minimal DMC energies
depends on the particular compound and on the specific structural
phase, but it is generally close to the value of 25\% typically used
within the hybrid density-functional theory.\cite{perdew1996b} Our
calculations provide yet another confirmation that the hybrid DFT indeed
provides an improved picture of investigated compounds compared to more
conventional functionals (LDA and GGA).


\section*{Acknowledgements}

 Support by NSF EAR-05301110, DMR-0804549 and OCI-0904794 grants
and by DOE DE-FG05-08OR23336
Endstation grant is gratefully acknowledged.
This study was enabled by INCITE and CNMS allocations at ORNL and by
allocation at NCSA. J.~K. would like to acknowledge financial support
by the Alexander von Humboldt Foundation during preparation of the
manuscript. We would like to thank M. Bajdich for numerous discussions
on wave function optimization.

\appendix\section{Cohesive energy of M\lowercase{n}O}
\label{sec:MnO_cohesion}

\begin{figure}
\resizebox{\linewidth}{!}{\includegraphics{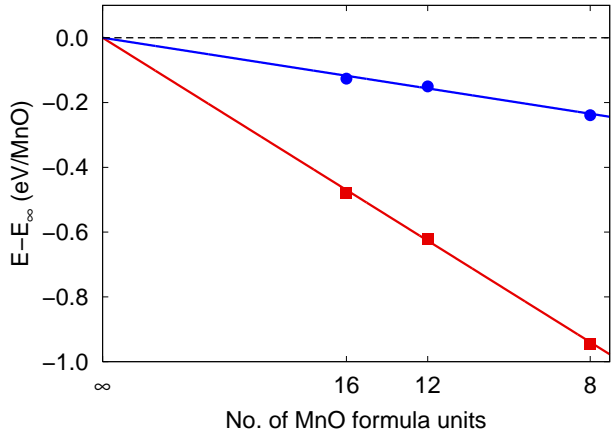}}
\caption{\label{fig:mno_fss}(color online) Finite-size scaling
  for MnO (B1 AFM-II phase at $V=21.7$ \AA${}^3$/MnO). All
  data points are averaged 
  over 8 $k$ points. Statistical error bars are smaller than symbol
  sizes. Red squares correspond to the Ewald formula for the Coulomb
  interaction 
  energy, blue points include additional correction described in the
  text. Lines are least-squares fits with
  $E(N)=a/N+E_{\infty}$.
}
\end{figure}

We have evaluated DMC energies of MnO crystal in three simulation
cells containing 16, 24 and 32 atoms. The results obtained with the
PBE0${}_{w=0.3}$ orbitals are plotted in
Fig.~\ref{fig:mno_fss}. Clearly, the averaging over 8 $k$ points
performed in all cases successfully suppresses differences in shapes
of the individual simulation cells and the remaining size dependence
can be very well fitted with a function $E(N)=a/N+E_{\infty}$. The
estimate for $E_{\infty}$ is $-120.203\pm 0.001$ hartree/MnO.

\begin{figure}
\resizebox{\linewidth}{!}{\includegraphics{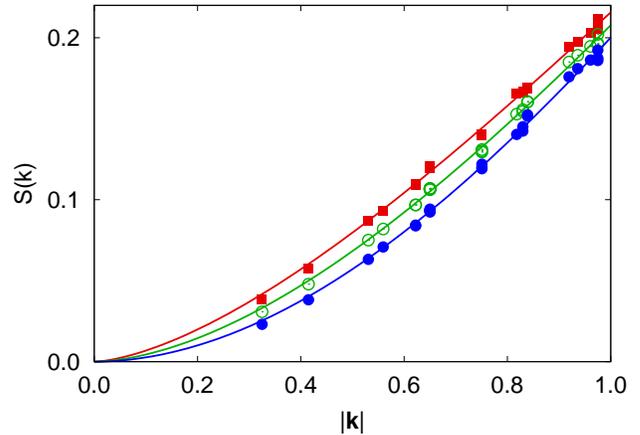}}
\caption{\label{fig:mno_sk}(color online) Static structure 
  factor $S(\mathbf{k})$ in MnO (B1 AFM-II phase at $V=21.7$
  \AA${}^3$/MnO). Calculations were performed with the PBE0${}_{w=0.3}$
  orbitals. Data are combined from simulation cells containing 16,
  24 and 32 atoms. Plotted are: VMC estimate (red squares), mixed DMC
  estimate (green empty circles) and extrapolated DMC estimate (blue
  filled circles). The lines are fits with
  $S(k)=\bigl[1-\exp(-\alpha k^{\beta})\bigr]$.}
\end{figure}

The $a/N$ term can be substantially reduced by techniques described in
Refs.~~\onlinecite{chiesa2006} and~~\onlinecite{drummond2008}.  One
part of the proposed correction to the Ewald total energy reads
\begin{equation}\label{eq:sk_corr}
\Delta E_{S(k)}=\frac1{4\pi^2}\int_D d^3\!k\,\frac{S(\mathbf{k})}{k^2}\,.
\end{equation}
The static structure factor $S(\mathbf{k})$ entering the expression is
defined as $S(\mathbf{k})=\langle\Psi_0|\hat\rho_{\mathbf{k}}
\hat\rho_{-\mathbf{k}}|\Psi_0\rangle/N$ with $\hat\rho_{\mathbf{k}}$
standing for a Fourier component of the electron density. The integral
in Eq.~\eqref{eq:sk_corr} runs over a domain $D$ centered around
$\mathbf{k}=0$ and having volume $8\pi^3/\Omega$, where $\Omega$ is
volume of the simulation cell. The structure factor $S(\mathbf{k})$ is
evaluated along the DMC simulation at a discrete set of points and
then extrapolated towards $\mathbf{k}=0$. We show this extrapolation
for the present case in Fig.~\ref{fig:mno_sk}, where we compare mixed
and extrapolated DMC estimates, Eqs.~\eqref{eq:mix_estim}
and~\eqref{eq:extrap_estim}. The small-$k$ behavior of $S(\mathbf{k})$
is found to be $\sim k^{1.7}$ for the mixed estimate and $\sim
k^{1.9}$ for the extrapolated one. This variance originates in a limited
quality of our trial wave functions at large distances between
electrons---the Jastrow factor we use is restricted to zero for
inter-electron separations larger than the Wigner--Seitz radius, and hence
the correct long-range asymptotics cannot be fully captured. In the end,
this deficiency is irrelevant, since the prediction of the extrapolated
estimate is sufficiently close to the exact asymptotics $\sim k^{2}$
that translates to $\Delta E_{S(k)}\sim
1/N$.\cite{chiesa2006,drummond2008}
The total energies corrected according to Eq.~\eqref{eq:sk_corr} are
shown in Fig.~\ref{fig:mno_fss} together with the pure Ewald data. The
correction reduces the finite-size errors by 75\%.

\begin{table}[b]
\caption{\label{tab:Mn_atom}Fixed-node DMC total energy of isolated
  Mn atom calculated using several sets of single-particle orbitals.}
\begin{tabular}{lc}
\hline\hline
single-particle orbitals  & $E$ (hartree) \\
\hline
restricted open-shell HF      & $-104.0133\pm 0.0006$ \\
unrestricted HF               & $-104.0185\pm 0.0005$ \\
unrestricted PBE0${}_{w=0.3}$  & $-104.0192\pm 0.0002$ \\
\hline\hline
\end{tabular}
\end{table}

For calculations of individual atoms we utilize the same form of the
trial wave function as we did in solids in order to stay within the
same level of theoretical description.  Single-particle orbitals for
atomic calculations were obtained using \textsc{Gamess}
code.\cite{gamess} Fixed-node DMC energies for the manganese atom
corresponding to several choices of orbitals are listed in
Tab.~\ref{tab:Mn_atom}. The difference between Hartree--Fock and
PBE0${}_{w=0.3}$ is negligible for our purposes, since it is smaller
than the statistical error bar achieved for the energy of the bulk
crystal.

Variation of single-particle orbitals in the trial wave function for the
oxygen atom does not lead to any appreciable differences of DMC
energies. To evaluate the cohesion of MnO, we use the result quoted in
Ref.~~\onlinecite{bajdich2008}, $E_O= -15.8421\pm 0.0002$ hartree.

\bibliography{qmc_papers,electronic_structure_methods,tmo,pseudopotentials}

\end{document}